\documentclass[a4paper,11pt]{article}
\usepackage{pos}
\usepackage{listings}
\usepackage{graphicx}
\title{Advancing Awkward Arrays for High-Performance CPU and GPU Processing}

\author*[a]{Ianna Osborne}
\author[b]{Manasvi Goyal}

\affiliation[a]{Princeton University,\\
Princeton, NJ, USA}

\affiliation[b]{Harvard University,\\
Cambridge, MA, USA}

\emailAdd{iosborne@princeton.edu}

\abstract{
Awkward Array is a Python library for representing and processing nested,
variable-length data that is widely used in high-energy physics. As HL-LHC analyses increasingly rely on accelerator hardware, efficient execution of irregular workloads has become essential. While dense numerical arrays map naturally to GPUs, nested and variable-length data structures remain significantly more difficult to accelerate because they require indirect indexing, segmented operations, and irregular memory access patterns. We present recent developments in the Awkward Array GPU backend, including CUDA implementations built on NVIDIA CUDA Core Compute Libraries (CCCL), optimized memory management, and segmented reduction algorithms for ragged arrays. These
developments preserve the existing Python programming model while
substantially improving GPU throughput on irregular workloads. We describe the
backend architecture, automated validation framework, and benchmark results
comparing CPU, CuPy, and CUDA implementations.
}

\FullConference{
23rd International Workshop on Advanced Computing and Analysis Techniques in Physics Research (ACAT2025)\\
8--12 September 2025\\
Hamburg, Germany
}

\begin{document}
\maketitle

\section{Introduction}

The High-Luminosity Large Hadron Collider (HL-LHC)~\cite{hl-lhc} will significantly increase
the volume and complexity of data available for physics analyses. Efficient
processing of irregular event structures has therefore become increasingly
important for modern high-energy physics workflows.

Awkward Array provides a columnar representation for nested and variable-length
data while maintaining a NumPy-like interface~\cite{awkward}. It has become a key component of Python-based analysis ecosystems, enabling efficient manipulation of detector and event data. Unlike dense multidimensional arrays, irregular data structures require indirect indexing, variable-length segments, and dynamic memory layouts, making efficient parallel execution challenging. As a result, many optimization techniques developed for dense numerical arrays do not directly apply.

This work presents recent developments toward high-performance execution of Awkward Array across CPU and GPU backends, including CUDA~\cite{cuda} implementations of irregular operations based on NVIDIA CUDA Core Compute Libraries (CCCL)~\cite{cccl}, optimized GPU memory management, an automated validation framework for ensuring backend consistency, and performance evaluation on representative irregular workloads.

\section{Backend Architecture}

\begin{figure}[ht]
\centering
\includegraphics[height=3.5cm]{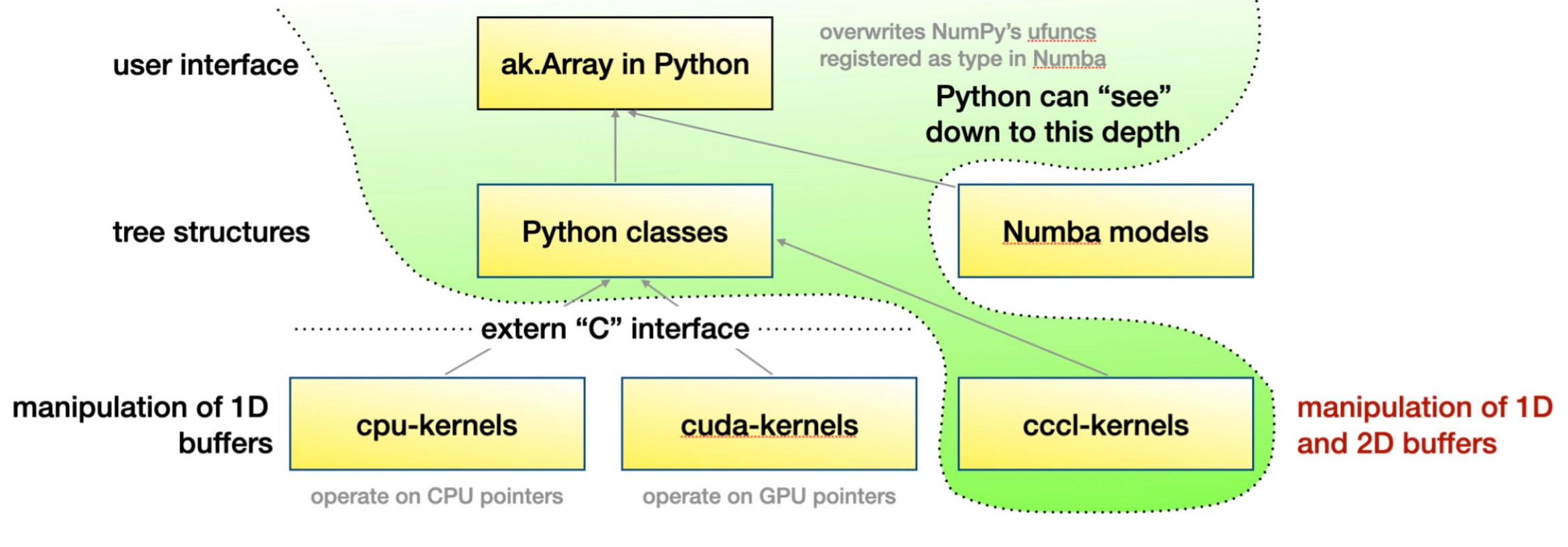}
\vspace{-4mm}
\caption{Awkward Array execution model. High-level Python operations are
dispatched to backend-specific CPU or CUDA implementations.}
\label{fig:architecture}
\end{figure}

Awkward Array separates high-level operations from backend-specific kernel
implementations. This layered architecture allows users to execute identical
analysis code on either CPU or GPU backends. For example, switching execution from the CPU to the CUDA backend requires changing only the backend argument, with no modifications to the analysis code. Backend implementations expose the same high-level API while dispatching execution to backend-specific kernel libraries. This separation allows new hardware backends to be developed independently without modifying user analysis code. The CPU backend currently consists of approximately 144 specialized kernels. Many high-level operations map directly to a single CPU kernel implementation. The CUDA backend, however, relies on CuPy for memory management and kernel launching~\cite{cupy} while maintaining the same high-level Python API. The CPU backend serves as the correctness reference and shares the same high-level execution model, illustrated in Figure~\ref{fig:architecture}.

\section{GPU Acceleration with CUDA and CCCL}

Recent development efforts focus on improving GPU performance through optimized memory management, increased parallelism, and integration with NVIDIA CUDA Core Compute Libraries (CCCL)~\cite{cccl}. GPU execution of irregular reductions requires several preprocessing stages, including segment construction, temporary workspace allocation on the GPU, and invocation of segmented reduction primitives. Consequently, a single logical Awkward operation may require multiple CUDA kernel launches. Many GPU operations therefore follow a common execution pipeline that constructs segment offsets, allocates output and temporary storage, invokes CuPy kernels or CCCL segmented primitives, and writes the resulting values back into Awkward layouts.

\begin{figure}[ht]
\vspace{-9mm}
\centering
\begin{minipage}[c]{0.6\textwidth}
    \centering
    \includegraphics[height=6.0cm]{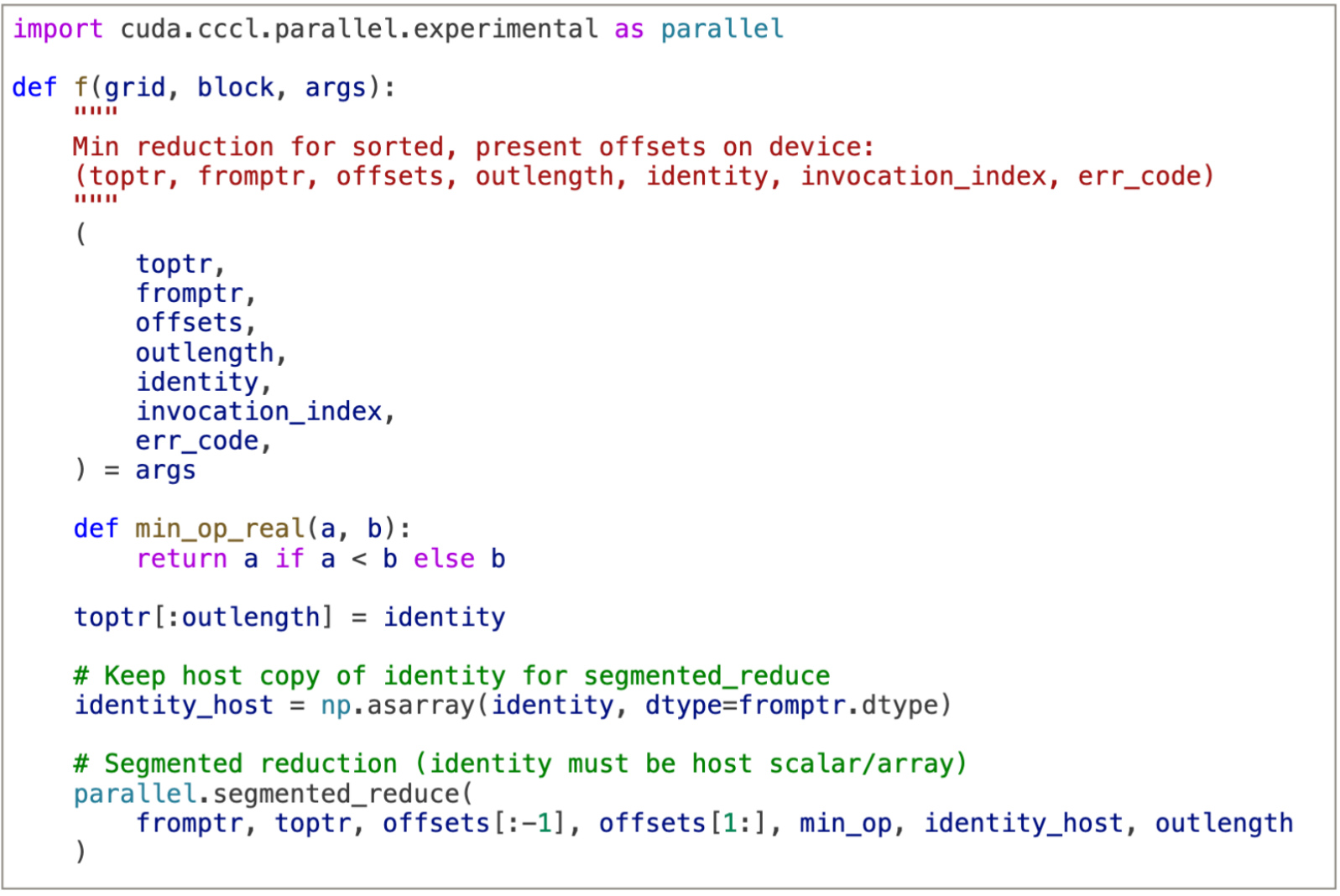}
\end{minipage}%
\hspace{-0.08\textwidth}%
\begin{minipage}[c]{0.40\textwidth}
    \centering
    \includegraphics[height=7.84cm]{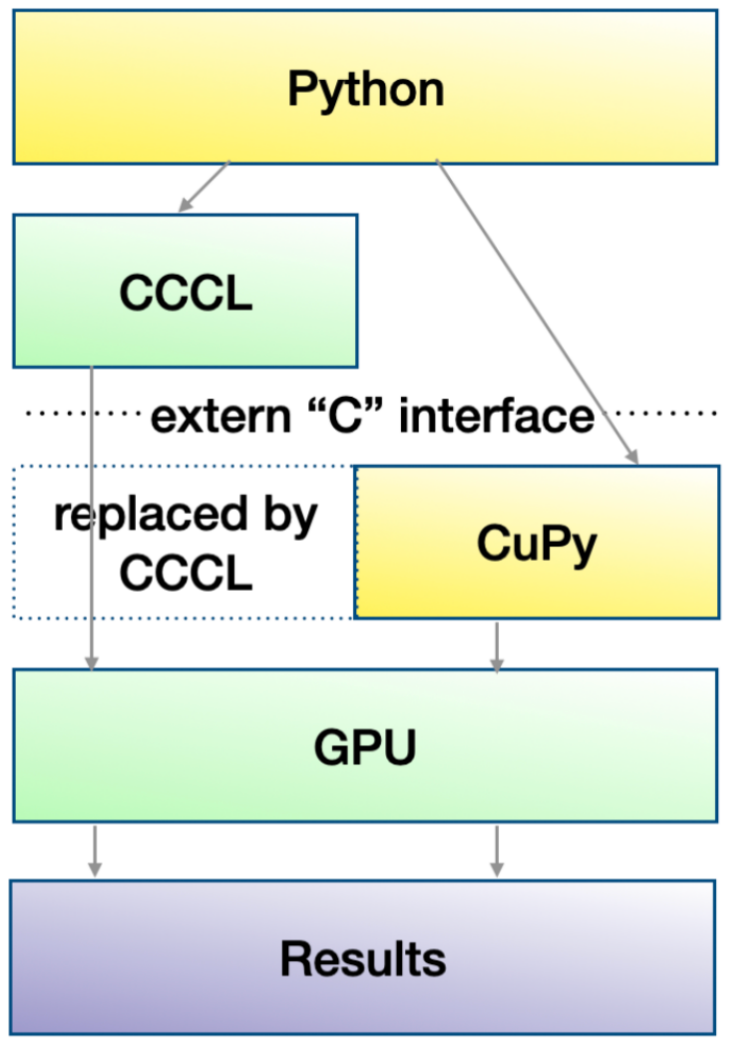}
\end{minipage}
\vspace{-8mm}
\caption{Current integration of CCCL within the GPU backend. The CCCL implementation simplifies the segmented reduction code (left), while multiple software layers in the current backend contribute to kernel-launch overhead (right).}
\label{fig:cccl-cuda}
\end{figure}

CCCL provides segmented GPU primitives that are well suited for these operations~\cite{cccl,cub,thrust}. For example, segmented reductions compute independent reductions over each ragged list while efficiently exploiting GPU parallelism. Replacing custom CUDA kernels with CCCL primitives substantially simplifies the implementation while preserving the existing Python interface. For the segmented minimum operation, approximately 70 lines of custom CuPy RawKernel code are replaced by a three-line Python call to a CCCL segmented reduction primitive illustrated in Figure~\ref{fig:cccl-cuda}. The work presented here focuses on segmented reduction operations, including minimum, maximum, sum, product, and count. Together, these primitives provide the foundation for accelerating a broad class of Awkward Array computations.

\section{Validation and Performance Evaluation}

Maintaining identical behavior across CPU and GPU implementations requires extensive validation. Kernel specifications are shared across backends, allowing identical correctness tests to be generated for both CPU and CUDA implementations. The validation framework combines kernel-level unit tests, backend consistency checks, integration tests for high-level Awkward operations, and regression tests for newly introduced CUDA kernels. Benchmarks were performed on a system equipped with an AMD EPYC 7H12 processor (128 physical cores) and an NVIDIA A100 PCIe 40\,GB GPU. The benchmark workload evaluates segmented reductions over irregular arrays with increasing numbers of variable-length segments. Each benchmark was executed multiple times after GPU warm-up, and the reported runtime corresponds to the minimum execution time.

\begin{figure}[ht]
\centering
\begin{minipage}{0.49\textwidth}
    \centering
    \includegraphics[width=\linewidth]{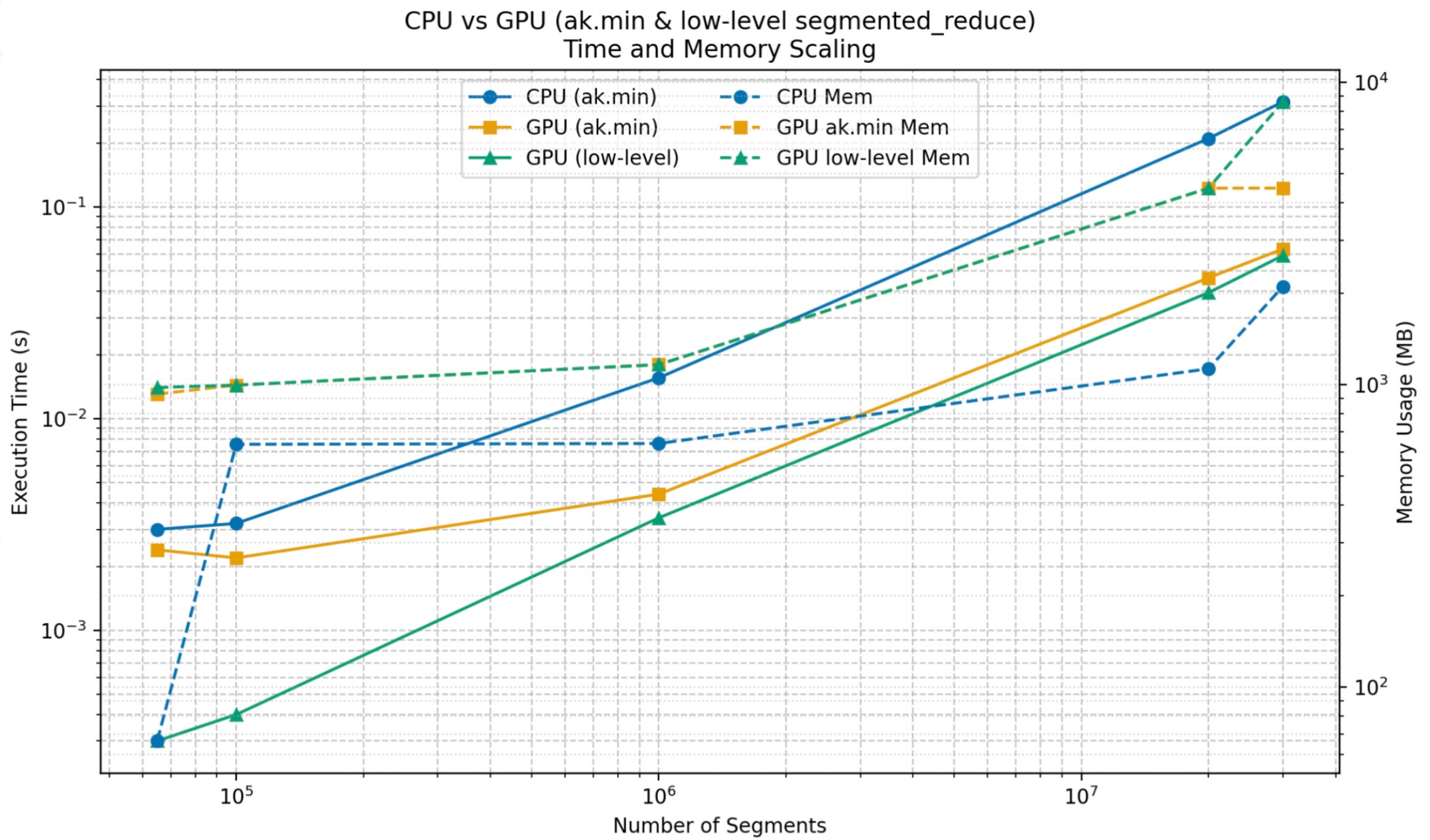}
\end{minipage}\hfill
\begin{minipage}{0.49\textwidth}
    \centering
    \includegraphics[width=\linewidth]{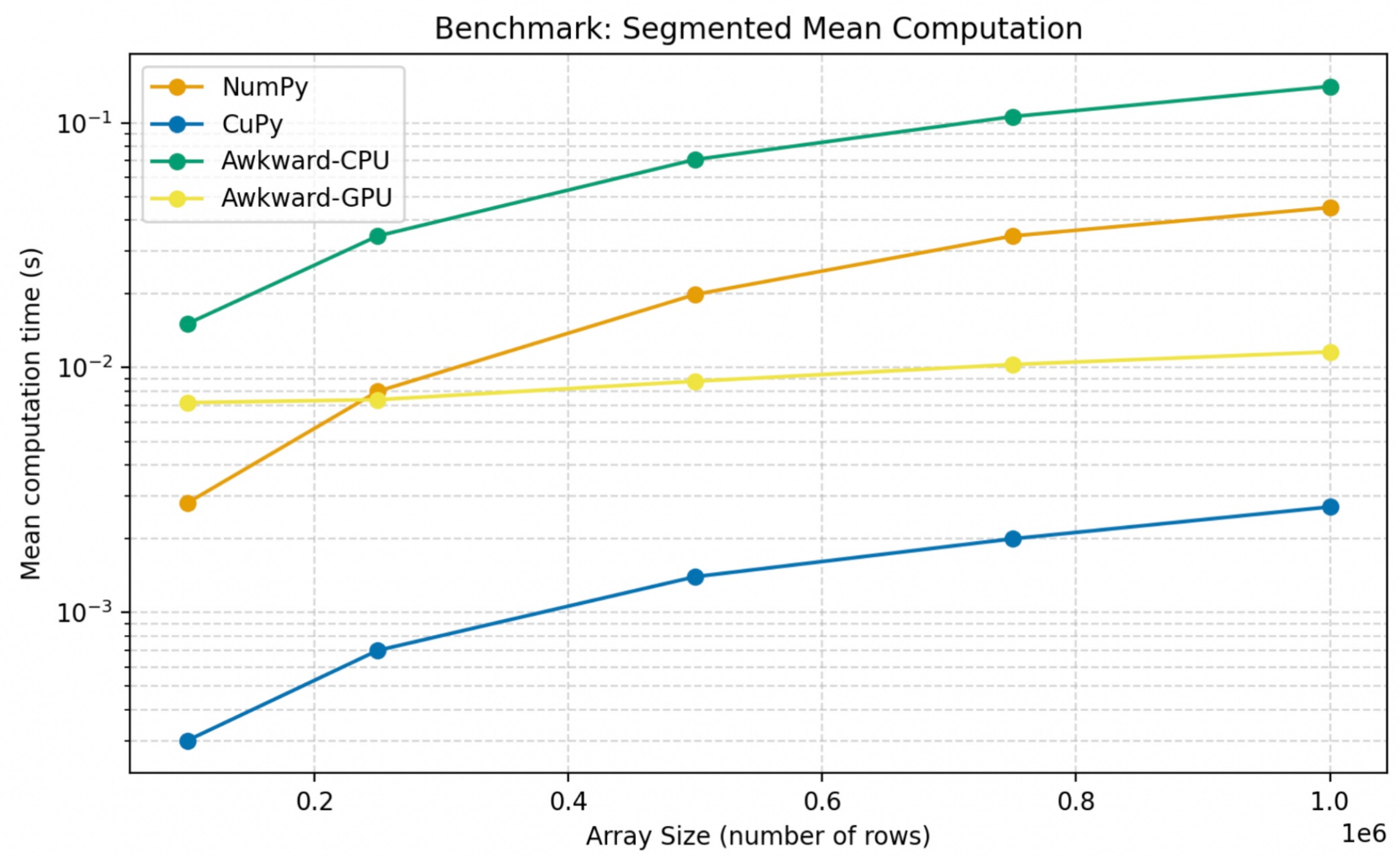}
\end{minipage}
\caption{Execution time and peak memory usage for the Awkward \texttt{ak.min} implementation and the underlying CCCL segmented reduction (left), and segmented mean performance comparing NumPy, CuPy, and the Awkward CPU and GPU backends (right). At the largest workload, GPU \texttt{ak.min} is 5.3$\times$ faster than the CPU implementation, while the low-level CCCL primitive achieves nearly 6$\times$ speedup.}
\label{fig:runtime-memory}
\end{figure}

Figure~\ref{fig:runtime-memory} (left) shows both execution time and peak memory usage for \texttt{ak.min} and the underlying CCCL segmented reduction. At the largest workload, GPU \texttt{ak.min} achieves a 5.3$\times$ speedup over the CPU implementation, while the low-level CCCL primitive reaches nearly 6$\times$ speedup, indicating that most remaining overhead arises from the high-level Awkward execution pipeline. Peak GPU memory usage increases from approximately 1\,GB to 4\,GB across the benchmark range, compared with roughly 60\,MB to 2\,GB for the CPU backend, primarily owing to temporary storage required for offset construction and segmented reduction operations. The segmented mean benchmark (right) compares NumPy~\cite{numpy}, CuPy~\cite{cupy}, and the Awkward CPU and GPU backends. The GPU implementation provides up to 12$\times$ speedup over the Awkward CPU backend and approximately 4$\times$ speedup over NumPy for the largest benchmark. Although it remains around 4$\times$ slower than a hand-written CuPy implementation, the performance gap narrows with increasing problem size, indicating that the remaining overhead is largely attributable to the generality of the high-level Awkward execution model rather than the underlying GPU kernels.

\section{Conclusions}

Awkward Array provides an efficient framework for processing irregular scientific data across both CPU and GPU architectures. Recent developments based on CUDA and CCCL substantially improve GPU throughput while preserving the familiar Python programming model. The new GPU backend combines optimized memory management with high-performance segmented primitives for irregular data, enabling significant speedups over CPU implementations on large workloads. Several challenges remain, including reducing kernel-launch overhead introduced by the current execution pipeline and adapting to the evolving CCCL API. These developments establish the foundation for compiler-driven optimization of high-level Awkward Array expressions, including kernel fusion and portable execution across future accelerator platforms.

\acknowledgments

This work was supported by the National Science Foundation under Cooperative Agreements OAC-1836650, PHY-2323298 and PHY-2121686.

\end{document}